# Emerging magnetism and anomalous Hall effect in iridate–manganite heterostructures

John Nichols[1], Xiang Gao[1], Shinbuhm Lee[1], Tricia L. Meyer[1], John W. Freeland[2], Valeria Lauter[3], Di Yi[4], Jian Liu[5], Daniel Haskel[2], Jonathan R. Petrie[1], Er-Jia Guo[3], Andreas Herklotz[1], Dongkyu Lee[1], Thomas Z. Ward[1], Gyula Eres[1], Michael R. Fitzsimmons[3], and Ho Nyung Lee[1]*

[1]Materials Science and Technology Division, Oak Ridge National Laboratory, Oak Ridge, TN, 37831, USA

[2]Advanced Photon Source, Argonne National Laboratory, Argonne, IL, 60439, USA

[3]Quantum Condensed Matter Division, Oak Ridge National Laboratory, Oak Ridge, TN, 37831, USA

[4]Department of Applied Physics, Stanford University, Stanford, California, 94305, USA

[5]Department of Physics and Astronomy, University of Tennessee, Knoxville, TN, 37996, USA

*Correspondence to: hnlee@ornl.gov

Strong Coulomb repulsion and spin-orbit coupling are known to give rise to exotic physical phenomena in transition metal oxides. Initial attempts to investigate systems where both of these fundamental interactions are comparably strong, such as 3$d$ and 5$d$ complex oxide superlattices, have revealed properties that only slightly differ from the bulk ones of the constituent materials. Here, we observe that the interfacial coupling between the 3$d$ antiferromagnetic insulator SrMnO$_3$ and the 5$d$ paramagnetic metal SrIrO$_3$ is enormously strong, yielding an anomalous Hall response as the result of charge transfer driven interfacial ferromagnetism. These findings show that low dimensional spin-orbit entangled 3$d$-5$d$ interfaces provide an avenue to uncover technologically relevant physical phenomena unattainable in bulk materials.

The strong interplay among charge, spin, orbital, and lattice order parameters in transition metal oxides (TMOs) is known to produce exotic physical phenomena[1], which can be significantly tuned through interfacial coupling between dissimilar materials[2]. Examples include enhanced superconducting critical temperature in cuprate bilayers[3], formation of a two-dimensional electron gas at an interface between two band insulators[4], improved transport and thermoelectric properties by fractional control of interfacial composition[5,6], and conducting interfaces between transparent titanates[7]. Although there have been several studies of interfacial magnetism in manganite[8-12] and ferrite[13] superlattices, they exclusively involve 3$d$ and 4$d$ TMOs. Even though there are a few examples of successful synthesis of 3$d$-5$d$ superlattices[14-17], there are no examples of strong interfacial coupling between these materials as the field remains in its infancy. With the emergence of a novel insulating ground state with effective total angular momentum $J_{eff}$ = 1/2 that is induced by strong spin-orbit coupling (SOC), there has been enormous interest in many Ir-based 5$d$-TMOs[18-21], which have a SOC interaction strength ($\xi$) with an energy scale comparable to the on-site Coulomb interaction ($U$)[22]. This interest is largely due to theoretical predictions of exotic physical properties such as unconventional superconductivity [23], Weyl semi-metals[20], and topological insulators[24,25] in 5$d$ systems. However, these novel ground states are yet to be experimentally confirmed. In order to narrow this gap between experimental and theoretical efforts, we have synthesized atomic-scale heterostructures by incorporating the antiferromagnetic insulator SrMnO$_3$ (SMO), a 3$d$ TMO with weak $\xi$ (0.01 – 0.1 eV) strong $U$ (5 – 7 eV), and the paramagnetic metal SrIrO$_3$ (SIO), a 5$d$ TMO with strong $\xi$ (0.1 – 1 eV) and modest $U$ (1 – 3 eV)[22]. Such a sample geometry uniquely enables the investigation of 3$d$-5$d$ interfaces where collectively both $U$ and $\xi$ are stronger than in either parent compound. Interestingly, we find that our [(SMO)$_m$/(SIO)$_n$]$_z$ (M$_m$I$_n$) heterostructures, where *m* and *n* are, respectively, the thicknesses of SMO and SIO in unit cells, display exceptionally strong interfacial coupling between the two constituent materials, yielding a ferromagnetic ground state. Such emerging interfacial magnetism in turn results in a strong anomalous Hall effect (AHE). As the emergence of ferromagnetism and the anomalous Hall effect are completely absent from either parent compound, this discovery provides the first experimental evidence of strong coupling at the interface of 3*d* and 5*d* materials.

**Results**

**Emerging magnetism**

The first indication of such unique behavior is the onset of magnetism in atomically thin superlattices. The macroscopic magnetic properties were measured with a superconducting quantum interference device (SQUID) and are shown in Fig. 1. The magnetic field ($H$) dependence of symmetric samples ($m = n$) is presented in Fig. 1 and clearly reveals the fact that samples with the thinnest layers (i.e. atomically thin superlattices) have the largest magnetic response. Although this is certainly a ferromagnetic response emerging at the SIO/SMO interface, the facts that the overall magnetization ($M$) of $M_1I_1$ is significantly larger than twice that of $M_2I_2$ along with $M_4I_4$ having $M \approx 0$ implies that interfacial diffusion is not responsible for the magnetic properties here and the mechanism driving this induced interfacial magnetism must have a characteristic length scale of just a few unit cells. The temperature dependent nature of the magnetization of these samples is shown in Fig. 1b. Consistent with the field sweeps, all samples with $m > 3$ showed no magnetic order, while below this limit, the magnitude increased with decreasing $m$. The Curie temperature ($T_c$) is shown in the inset where $M_1I_1$ has the largest $T_c$

~ 190 K. Note that $M_1I_1$ has a second anomaly at ~120 K for $H \perp c$ that is likely associated with its electronic properties as discussed below. The magnetic anisotropy of a second $M_1I_1$ sample is presented in Fig. 1c and 1d. Note that although the saturation magnetization and $T_c$ are independent of the direction of $H$, both the coercive field and remnant magnetization are roughly an order of magnitude larger when $H$ is parallel to the $c$-axis (out-of-plane). This result implies that the $c$-axis is the magnetic easy axis.

**Elemental specific characterization by x-rays and neutrons**

In order to fully understand the magnetism of these superlattices, it is necessary to identify the relative contribution of Mn and Ir ions to the overall magnetic moment. Both x-ray absorption (XAS) and x-ray magnetic circular dichroism (XMCD) spectra provide information rich with elemental specific contributions regarding both the electronic and magnetic structures. Thus, we collected XAS and XMCD spectra near the $L_3$ and $L_2$ edges of both Mn and Ir (Fig. 2) in order to understand the underlying mechanism of the novel magnetism. The XAS peak position of the Mn $L_3$ edge show that the onset of magnetism is accompanied by a shift of this peak to a lower energy, which implies that the Mn oxidation state in the heterostructures are lower than $Mn^{4+}$ found in stoichiometric SMO. Similarly, the position of the Ir $L_3$ edge shifts to a higher energy and indicates that the Ir oxidation state are enhanced relative to $Ir^{4+}$ of stoichiometric SIO. It is important to note that even if the oxidation state of the constituent materials deviates from their nominal values, our data still convincingly indicates that in order to maintain charge balance, there is charge transfer from the SIO to the SMO layers resulting in electron (hole) doped SMO (SIO) layers. The average oxidation states are estimated from the peak shifts and are presented in the inset of Fig 2, where $M_1I_1$ clearly has the largest deviation from the nominal value with a charge transfer of ~0.5 electron/hole per perovskite unit cell. Although in absolute units these estimates of the oxidation state have a relatively large uncertainty, it is important to note that their relative uncertainties are significantly smaller than the data points. The XMCD spectra of the Mn $L_3$ edge show that $M_1I_1$ has a large negative response, which indicates that the magnetic moment of the Mn ions ($M_{Mn}$) orders parallel to $H$. As $m$ increases, the Mn XMCD decreases. Despite the consistency between SQUID and Mn XMCD measurements, there are surprisingly finite XMCD peaks near the Ir $L$ edges. This XMCD result implies that there is a net magnetic moment of Ir ($M_{Ir}$) due to the onset of ferromagnetism or canted antiferromagnetism. The observation of net ferromagnetic order of Ir ions is quite surprising since $Ir^{4+}$ and $Ir^{5+}$ tend to favor antiferromagnetic[26,27] and paramagnetic[28,29] ground states, respectively. Thus, varying the valence state of Ir may provide a phase diagram as rich as those already known to the manganites. We were able to apply sum rules to the Ir XMCD spectra in order to separate the spin ($S$) and orbital ($L$) contributions of $M_{Ir}$ and found them to be 0.013 $\mu_B$ and 0.057 $\mu_B$, respectively for $M_1I_1$, whereas for $M_{Mn}$ $S$ and $L$ are 0.9 $\mu_B$ and 0.3 $\mu_B$, respectively. Combining these results, we determine the total magnetization ($M = L + 2S$) in each material to be $M_{Ir} = -0.08$ $\mu_B$/Ir and $M_{Mn} = 2.1$ $\mu_B$/Mn, which are in good agreement with SQUID data. Thus, we conclude that $M_{Mn}$ is mostly driven by spin while $M_{Ir}$ has predominately orbital contributions due to strong SOC[19]. Additionally, the XAS branching ratio (BR = $I_{L3}/I_{L2}$) of Mn in SMO is ~2 and systematically increases with decreasing $m$. Although this qualitative behavior can be explained by the reduction of the Mn oxidation state, it is worth noting that a BR > 2 is often attributed to the presence of spin-orbit interactions[30].

The microscopic origin of the magnetism was further investigated by polarized neutron reflectometry (PNR), which is a sensitive probe of spin asymmetry. This technique provides a detailed look at the magnetism of thin films and heterostructures as a function of depth. However, PNR of our symmetric magnetic superlattices is a formidable task since only short period superlattices ($m \leq 3$) are magnetic and all superlattice Bragg peak positions of these samples lie at wavevector transfer ($q$) values unobtainable with reasonable measurement parameters. This challenge was overcome by synthesizing an asymmetric $M_1I_{10}$ sample, which has a larger superlattice period and an appreciable magnetic response (see Supplementary Figure 2b). As shown in Fig. 3a−c, we observed a finite spin asymmetry which is a clear indication of ferromagnetic order and, thus, $M_1I_{10}$ is also ferromagnetically ordered. The chemical and neutron scattering length density (SLD) profiles obtained from spin dependent PNR measurements and x-ray SLD profile from x-ray reflectometry (XRR) are shown in Fig. 3d. Note that although it is typical, the apparent broadness of the SLDs arise from there being 13 SLDs that all differ by less than two standard deviations (2-sigma) from the ideal fit, indicating that this model is extremely robust. From this PNR result, the magnetic depth profile is determined and presented in Fig. 3d. Notice that $M_{Mn}$ is much larger than $M_{Ir}$, which is consistent with XMCD measurements. However, conversely, our PNR indicates that $M_{Ir}$ aligns parallel to the applied magnetic field in $M_1I_{10}$, whereas XMCD has revealed that it aligns antiparallel for $M_1I_1$. This discrepancy suggests that there is a critical SIO thickness, in which the Ir moments realign. Confidence in this interpretation of non-zero $M_{Ir}$ arises from the fact that if the $M_{Ir}$ is forced to zero (dashed lines in Fig. 3a−c), the model significantly deviates from the experimental data. Moreover, if $M_{Ir}$ is forced to align antiparallel to $M_{Mn}$, similar to XMCD of $M_1I_1$, this separation is exacerbated (see Supplementary Figure 2a). The thickness averaged $M$ values for the Mn and Ir-layers obtained from PNR is in excellent agreement with that obtained from SQUID measurements (see Supplementary Figure 2b)—further evidence that in-plane components of $M_{Mn}$ and $M_{Ir}$ for the $M_1I_{10}$ sample are parallel. In addition, recall that bulk SIO is paramagnetic and, even though a small ferromagnetic response has also been observed in SIO under reduced dimensionality in other studies: $Sr_2IrO_4$[31,32] and $(STO)_1/(SIO)_n$ ($n \leq 3$)[15], our observation provides the first example of ferromagnetism in thick slabs of SIO which clearly arises from strong interfacial coupling between 3$d$ and 5$d$ TMOs.

**Transport properties and Hall measurements**

The electronic properties of the symmetric samples were investigated via $dc$-transport measurements, and the temperature dependent sheet resistance ($R_S$) is shown in Fig. 4a. SMO (data not shown) is too resistive to measure [$R_S$ (300K) ~ 1 MΩ] and SIO is semi-metallic. The fact that all samples are roughly 50 nm thick and the resistance of $M_{12}I_{12}$ is approximately double that of SIO implies that the SIO layers in long period superlattices ($m \gtrsim 12$) dominate the overall electronic conduction. However, when the layer thicknesses are intermediately thick ($3 \leq m \leq 6$), the heterostructures have significantly enhanced metallicity with a weak upturn below 50 K which is most probably due to weak localization. In this intermediate thickness region, there is minimal charge transfer which implies that the magnitude of electron (hole) doping of the SMO (SIO) layers is quite small. Since bulk SMO is known to be insensitive to small concentrations of electron doping[33], the enhanced metallicity observed in the intermediate-period superlattices likely resides within the SIO layers. This result also indicates that SIO is sensitive to small concentrations of hole doping. As the layer thickness is further reduced ($m < 3$), the resistance increases as shown in Fig. 4a. This is somewhat counterintuitive since one would expect the onset of ferromagnetism to coincide with the enhanced metallicity. Consider the resistivity of $M_1I_1$, which displays a semi-metallic behavior with a local maximum at ~120 K. Comparing this to comparably doped bulk $La_{1-x}Sr_xMnO_3$ ($x = 0.55$)[33], we observe a quantitatively similar temperature dependent resistivity behavior that is roughly an order of magnitude larger than our $M_1I_1$ superlattice. Thus, the resistivity in small period superlattices is explained by the large electron doping concentration in atomically thin SMO layers resulting in a finite electrical conductivity accompanied by the atomically thin SIO layers having reduced conductivity due to reduced dimensionality[18], a large concentration of hole dopants[34], or the finite thickness effect[35,36]. Therefore the enhanced metallicity in the intermediately thick samples is due the SIO layers while the finite conductivity for $M_1I_1$ is due to the onset of conductivity in the SMO layers.

The intriguing magnetic properties of these superlattices were further investigated via magnetoresistance (MR) and Hall measurements presented in Fig. 4c–f. The MR of the $M_1I_1$ sample (Fig. 4c) has a negligible response at room temperature. However, a negative linear response starts to appear near and below $T_c \sim 190$ K and increases systematically in magnitude with further decreasing temperature. Interestingly, at lower temperatures ($T < 75$ K), a butterfly hysteresis loop appears at small $H$ that is coupled to the coercive field as comparatively shown in Fig. 4c and 4e. Comparing the low temperature MR for different samples (see Fig. 4d) indicates that all superlattices have a negative MR response that increases in magnitude with the onset of magnetism, whereas the SIO film has a small positive response. These behaviors are consistent with typical results for ferromagnets and paramagnets, respectively. Strikingly, the Hall measurements ($R_{xy}$) of our superlattices lead to an unprecedented observation. Consider the temperature dependent Hall resistance of the $M_1I_1$ superlattice shown in Fig. 4e. Above $T_c$, the Hall response is linear with a negative slope, indicating $n$-type carriers. Below $T_c$ a nonlinear

AHE appears, opening a large hysteresis loop at low temperatures with a shape and coercive field practically identical to $M(H)$ sweeps obtained from SQUID measurements (see Supplementary Figure 4). In addition, it is evident from Fig. 4f that only the magnetic samples display the AHE response. Thus, it is indisputable that the AHE observed here is due to the onset of magnetism in this system.

**Discussion**

Since the dominant magnetic ion is Mn and the AHE is driven by magnetism, it is logical to assume the majority of AHE resides within the SMO layers. Recent advances in understanding the AHE separates such materials into three categories: (i) the dirty metal limit where intrinsic and side jump scattering leads to a scaling relationship between the transverse conductivity ($\sigma_{xy}$) and longitudinal conductivity ($\sigma_{xx}$) of $\sigma_{xy} \propto \sigma_{xx}^{1.6}$; (ii) the super clean metal limit where skew scattering off extrinsic defects leads to a scaling relationship of $\sigma_{xy} \propto \sigma_{xx}$; and (iii) the moderately dirty metal region where intrinsic scattering leads to $\sigma_{xy}$ being approximately independent of $\sigma_{xx}$[37,38]. The latter has been modeled theoretically utilizing Berry phases and Berry curvature to successfully bridge the dirty and superclean limits and has been successful in modeling 3$d$ TMO systems within the range $3{,}000 \leq \sigma_{xx} \leq 450{,}000$ $\Omega^{-1}$cm$^{-1}$ [39]. Considering the scaling plot presented in Fig. 4b, we find that both $M_1I_1$ and $M_2I_2$ have $\sigma_{xx} \sim 2{,}000$ $\Omega^{-1}$cm$^{-1}$, which should place them in the dirty metal limit. However, fits to the low temperature data clearly show a much weaker power law ($\sigma_{xy} \propto \sigma_{xx}^{\varphi}$) than $\varphi = 1.6$. Recall that although the AHE resides in the SMO layers, the XMCD hinted that SOC substantially influences the SMO layers. Since the magnitudes of $\sigma_{xx}$ that separate the three regions described above depend inversely on $\xi$, for 5$d$ materials the moderately dirty limit is roughly $45 \leq \sigma_{xx} \leq 6{,}800$ $\Omega^{-1}$cm$^{-1}$. This remarkable agreement with our experimental results strongly suggests that SOC is instrumental in defining the novel magnetic and electronic ground states of these 3d-5d TMO heterostructures and that they are near the moderately dirty limit which has a characteristic dissipationless AHE current[40]. Another observation is the magnitude of $\sigma_{xy}$ observed here is significantly lower than the theoretical intrinsic scattering limit of ~900 $\Omega^{-1}cm^{-1}$ proposed by the Thouless-Kohmoto-Nightingale-Nijs formulism[41], despite the fact that $\sigma_{xy}$ should scale with $\xi$. We attribute this discrepancy to the fact that although the AHE resides in the SMO layers, the SIO layers still conduct appreciably well and serve as a resistive short of the voltage leads during the Hall measurements, which greatly reduce the measured values of $\sigma_{xy}$.

In summary, we have observed interfacial ferromagnetism that led to an anomalous Hall effect in atomic scale SMO/SIO superlattices grown on STO by pulsed laser epitaxy. This discovery provides clear experimental evidence of strong interfacial coupling between 3d and 5d materials. Furthermore, we have shown that SOC plays an integral role in defining these unique ground states and that this appears to be the prototypical system for investigating interfacial coupling between strong $U$ and strong SOC, thus presenting an avenue for potential spintronics applications. In addition, despite the Mn ions being the dominant magnetic host, we observe that the spins in Ir also ferromagnetically order opening a field of investigating magnetism in multivalent Ir ions. We believe that this work will stimulate further theoretical and experimental studies that will lead to greater understanding of the role of SOC in such systems.

## Methods

**Sample synthesis and structural characterization**

The superlattice samples of $[(SrMnO_3)_m/(SrIrO_3)_n]_z$ were synthesized by pulsed laser epitaxy on atomically flat $TiO_2$-terminated (100) $SrTiO_3$ substrates utilizing a KrF eximer laser ($\lambda = 248$ nm) with laser fluence, substrate temperature, and oxygen partial pressure of 1.0 J/cm$^2$, 700 ºC, and 100 mTorr, respectively. The crystal structure, orientation, phase purity, and crystallinity of these superlattices were determined by x-ray diffraction and reflectivity measurements.

**Magnetic and electrical measurements**

The macroscopic magnetic properties were characterized with a 7 T Quantum Design MPMS3. The XAS and XMCD spectra near the Mn and Ir $L$ edges were collected on beamlines 4-ID-C and 4-ID-D, respectively, at the Advanced Photon Source of Argonne National Laboratory. For the Mn $L$ edge data, both electron and fluorescence yields were simultaneously monitored. The Ir $L$ edges data were collected with a grazing incidence geometry and the fluorescence detection mode. The PNR measurements were performed on the Magnetism Reflectometer (beamline BL-4A)[42] at the Spallation Neutron Source of Oak Ridge National Laboratory and the magnetic depth profile was determined from fitting the neutron spin asymmetry that utilized the chemical model obtain from XRR. The *dc*-transport measurements were performed with a 14 T Quantum Design PPMS with a home-built user bridge. Contacts were made to all superlattice layers by ultrasonic soldering of gold wires with indium solder in a Van der Pauw configuration.

## Data availability

The data that support the findings of this study are available from the corresponding author upon request.

**Acknowledgements**

We would like to thank Michael A. McGuire for his experimental assistance as well as Tae-Won Noh, Changhee Sohn, Soyeun Kim, and Jun Sung Kim for valuable discussions and comments. This work was supported by the U.S. Department of Energy (DOE), Office of Science (OS), Basic Energy Sciences (BES), Materials Sciences and Engineering Division [synthesis, physical property characterization, x-ray absorption spectroscopy (XAS), x-ray magnetic circular dichroism (XMCD) and polarized neutron reflectometry (PNR) data analysis] and by the Laboratory Directed Research and Development Program of Oak Ridge National Laboratory, managed by UT-Battelle, LLC, for the U. S. DOE (PNR data fitting). The research at ORNL's Spallation Neutron Source was sponsored by the Scientific User Facilities Division, BES, U.S. DOE (PNR). Use of the Advanced Photon Source, an Office of Science User Facility operated for the U.S. DOE, OS by Argonne National Laboratory, was supported by the U.S. DOE under Contract No. DE-AC02-06CH11357 (XAS/XMCD). JL was sponsored by the Science Alliance Joint Directed Research and Development Program at the University of Tennessee.

## Author Contribution

JN and SL performed the sample synthesis. JN, XG, JRP, TLM, TZW, GE, DL, and HNL conducted the structural and basic physical property characterizations. JN, TLM, JWF, DY, JL, and DH collected and analyzed the XAS and XMCD spectra. JN, EG, AH, VL, and MRF contributed to the PNR measurements and data analysis. HNL and JN designed the experiment and wrote the manuscript with inputs from all authors. HNL initiated the research and supervised the work.

## Additional Information

**Supplementary Information** accompanies this paper at

http://www.nature.com/naturecommunications

**Reprints and permission** information is available online

athttp://npg.nature.com/reprintsandpermissions/.

**Competing financial interests:** The authors declare no competing financial interests.

Correspondence and requests for materials should be addressed to hnlee@ornl.gov.

**Figures and Figure Legends**

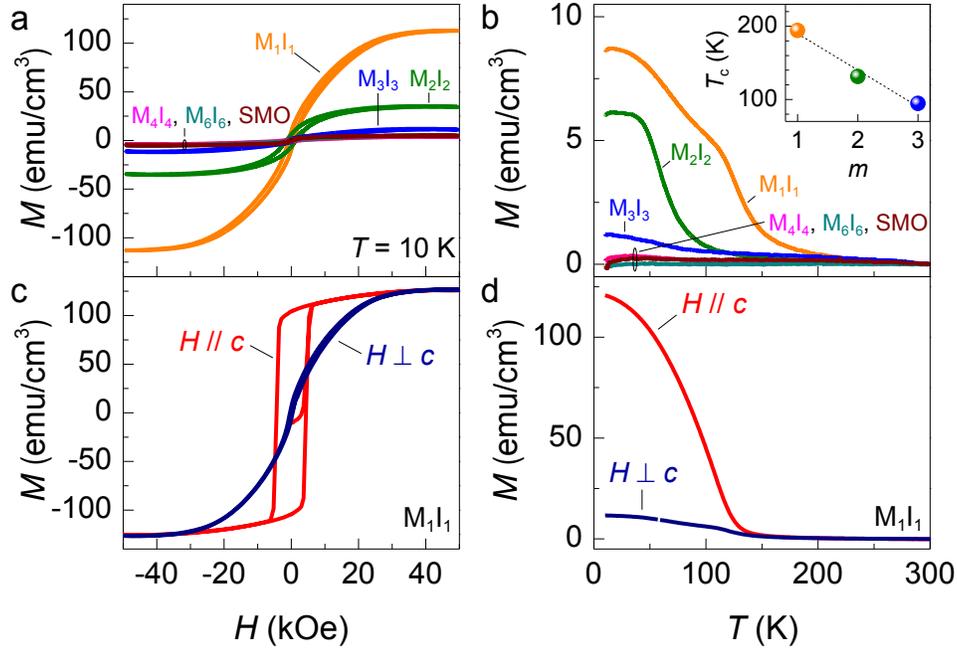

**Figure 1 Global Magnetization of SMO-SIO superlattices. a,** $M(H)$ of symmetric samples at $T = 10K$ after zero field cooling. **b,** $M(T)$ of symmetric samples at $H = 1$ kOe after field cooling in $H = 1$ kOe. The inset shows the SMO layer thickness ($m$) dependence of the Curie temperature. **c,** $M(H)$ of $M_1I_1$ at $T = 10K$ after zero field cooling. **d,** $M(T)$ of $M_1I_1$ at $H = 1$ kOe after field cooling in $H = 1$ kOe.

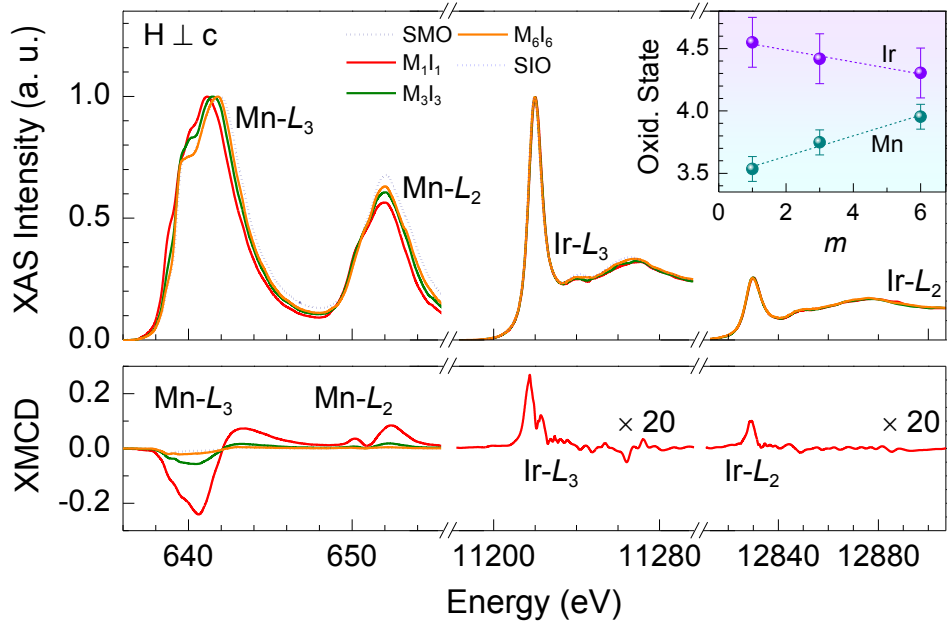

**Figure 2 Elemental specific charge transfer and interfacial magnetism by XAS and XMCD.** The data near the Mn (Ir) edges were obtained at $H$ = 50 kOe (40 kOe) with $H \perp c$ after cooling in zero field to 15K (10K). Both ions display a finite XMCD signal, which indicates that both SMO and SIO are ferromagnetically active. The peak near the $L_3$ edge of Mn (Ir) for the $M_1I_1$ sample shifts to lower (higher) energy indicating charge transfer from the SIO to the SMO layer. The inset shows the estimate of the oxidation state for each cation determined by a linear interpolation between known positions of Mn and Ir oxidation states, where the uncertainties were determined by propagating the instrumental energy uncertainties into oxidation state estimates.

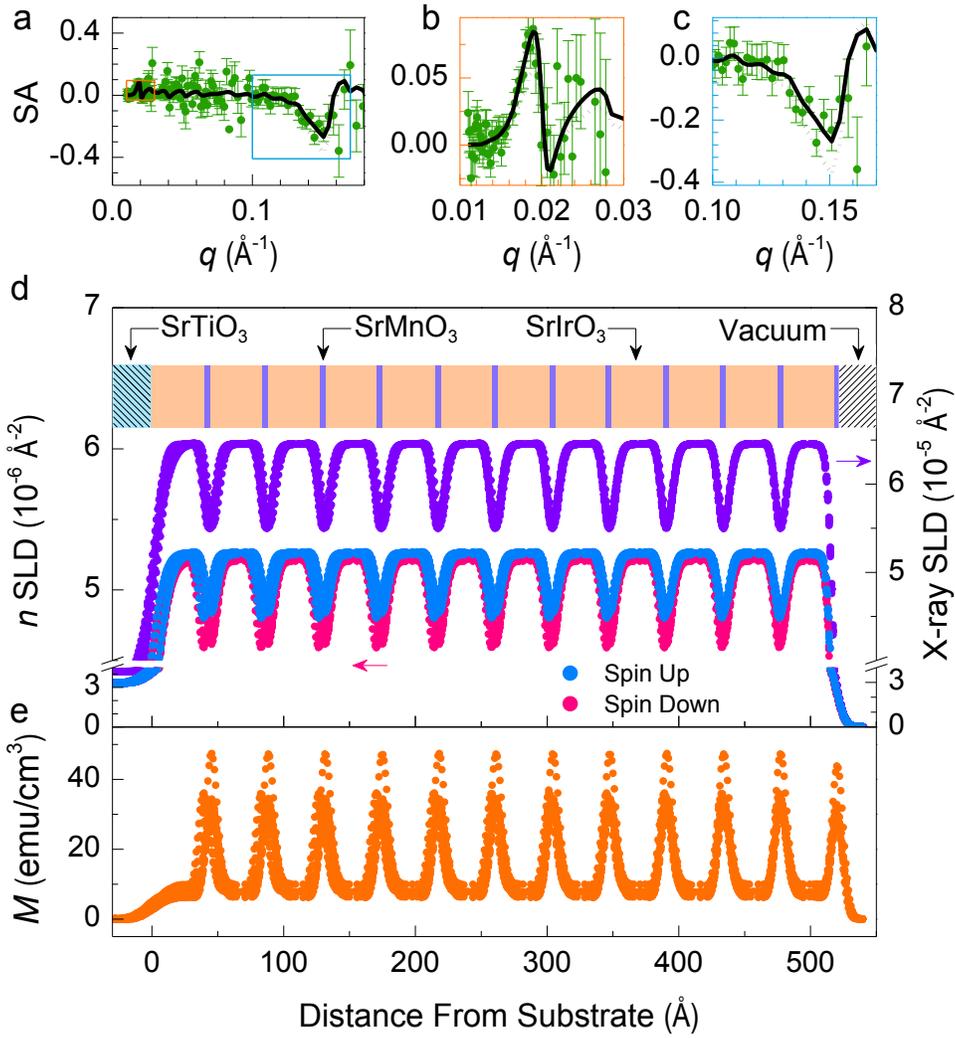

**Figure 3 Magnetic depth profiling by PNR.** Data was obtained from a [(SrMnO$_3$)$_1$/(SrIrO$_3$)$_{10}$]$_{12}$ superlattice on STO after a zero field cool at $T$ = 10 K and $H$ = 11.5 kOe with $H \perp c$. **a,** The spin asymmetry (SA = [R$_\uparrow$ - R$_\downarrow$]/[R$_\uparrow$ + R$_\downarrow$]), where solid (dotted) black lines represent models where the magnetism in the SIO layer is allowed to vary (forced to zero) for the fit. The orange and cyan rectangles represent the positions near the critical angle and first superlattice Bragg reflection shown in **b** and **c**, respectively. **d,** Depth profile of x-ray (purple) and neutron (blue and pink) scattering length densities, where a schematic drawing of the sample geometry is shown above the data. **e,** Magnetic depth profile obtained with fit parameters of $M_{Mn}$ = 85 emu/cm$^3$ (0.54 $\mu_B$/Mn) and $M_{Ir}$ = 9 emu/cm$^3$ (0.06 $\mu_B$/Ir).

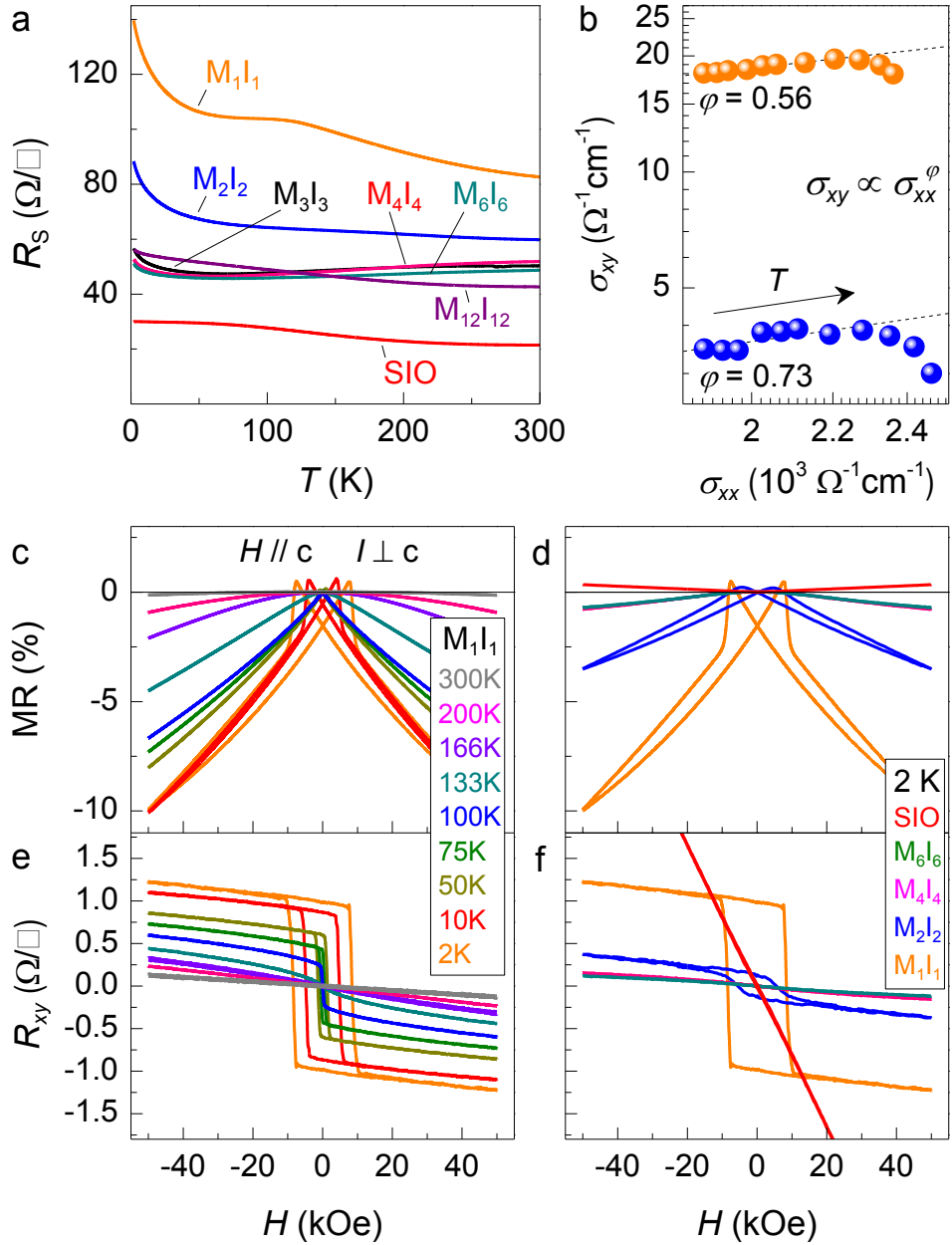

**Figure 4** *DC-transport and anomalous Hall effect.* **a,** $R_S(T)$ for SMO-SIO superlattices with all samples displaying semi-metallic and metallic behaviors. $MR(H) = [R_S(H) - R_S(0)]/R_S(0) \times 100\%$ with $H // c$ for **c** $M_1I_1$ at various temperatures and **d** short period samples ($n \leq 6$) at 2 K where the color scheme is identical to **a**. $R_{xy}(H)$ with $H // c$ for **e** $M_1I_1$ at various temperatures and **f** short period samples at 2 K that clearly display a nonlinear behavior attributed to a magnetism induced anomalous Hall effect. **b,** Scaling plot of $\sigma_{xy}$ and $\sigma_{xx}$ where they are determined using the total superlattice thickness.